\documentstyle[12pt,epsf]{article}
\newcommand{\beq}{\begin{equation}}
\newcommand{\eeq}{\end{equation}}
\newcommand{\beqa}{\begin{eqnarray}}
\newcommand{\eeqa}{\end{eqnarray}}

\newcommand{\qp}{\tilde{Q}_+}
\newcommand{\qm}{\tilde{Q}_-}

\newcommand{\pn}{M_{\pi^0}}

\newcommand{\vs}{\vspace{-0.25cm}}

\begin{document}

\begin{flushright}
{\small
UWThPh-1999-48 }
\end{flushright}


\vspace{2cm}

\begin{center}

{{\Large\bf
Virtual photons to fourth order CHPT with nucleons \footnote{Work
supported in part by TMR, EC-Contract No. ERBFMRX-CT980169 (EURODA$\Phi$NE).}
}}

\end{center}

\vspace{.4in}

\begin{center}
{\large
Guido M\"uller$^\star$\footnote{email:
gmueller@doppler.thp.univie.ac.at}}

\bigskip

$^\star${\it Universit\"at Wien, Institut f\"ur Theoretische Physik\\
Boltzmanngasse 5, A--1090 Wien, Austria}

\end{center}

\vspace{.7in}

\thispagestyle{empty}


Baryon chiral perturbation theory offers another possibility of investigating
isospin violation. As first stressed by Weinberg~\cite{weinmass}, reactions
involving nucleons and neutral pions can lead to gross violations of isospin,
e.g. in the scattering length difference $a(\pi^0 p) - a(\pi^0 n)$
he predicted an effect of the order of 30\%. This is
because chiral symmetry and isospin breaking appear  at the same order
and the leading isospin symmetric terms involving neutral pions are suppressed
due to chiral symmetry.  It is, however, known
that precise and complete one loop calculations in the baryon sector should
be carried out to fourth order since it has also been shown that
in many cases one loop graphs with exactly one dimension two insertion are
fairly large.
Most calculations in baryon CHPT are performed in  
heavy baryon chiral perturbation theory (HBCHPT) \cite{jm,bkkm}.
This is based on the observation
that a straightforward extension of CHPT with baryons treated fully
relativistically
leads to a considerable complication in the power counting since only
nucleon three--momenta are small compared to typical hadronic scales,
as discussed in detail in ref.\cite{gss}.
 However, one has to be careful with strict non--relativistic
expansions since in some cases they can conflict structures from analyticity,
as discussed e.g. in ref.\cite{bkmff}. Therefore, in ref.\cite{BL}, a novel
scheme was proposed which is based on the relativistic formulation but
circumvents the power counting problems (to one loop)
by a clever separation of the loop integrals into IR singular and regular
parts.
In this formulation, all analytic constraints are fulfilled by construction.

The starting point of our approach is to construct the most general
chiral invariant Lagrangian built from pions, nucleons and external
scalar, pseudoscalar, vector, axial--vector sources {\it and}
virtual photons, parametrized in terms of the vector field $A_\mu
(x)$. The Goldstone
bosons are collected in a 2$\times$2 matrix-valued field $U(x)= u^2(x)$.
The nucleons are described by structureless relativistic spin-${\small
  \frac{1}{2}}$ particles, the spinors being denoted by $\Psi
(x)$ in the relativistic case
or by the so--called light component $N(x)$ in the heavy fermion
formulation. The effective field theory admits a low energy expansion, i.e. the
corresponding effective Lagrangian can be written as
\beq
{\cal L}_{\rm eff} =
 {\cal L}_{\pi\pi}^{(2)} +  {\cal L}_{\pi\pi}^{(4)} +
 {\cal L}_{\pi N}^{(1)} +  {\cal L}_{\pi N}^{(2)} +
 {\cal L}_{\pi N}^{(3)} +  {\cal L}_{\pi N}^{(4)} + \ldots~,
\eeq
where the ellipsis denotes terms of higher order not considered
here. For the explicit
form of the meson Lagrangian and the dimension one and two
pion--nucleon terms, we refer to ref.\cite{bkmrev}. More precisely,
in the pion--nucleon sector,
the inclusion of the virtual photons modifies the leading
term of dimension one and leads to new
local (contact) terms starting at second order \cite{ms}. In
particular, since the electric charge related to the virtual photons
always appears quadratic,
the following pattern for the terms in the electromagnetic effective
Lagrangian emerges.
At second order, we can only have terms of order
$e^2$, at third order $e^2 q$ and at fourth order $e^2 q^2$ or $e^4$
(besides the standard strong terms).The inclusion of the virtual 
photons proceeds  with,
\beq
Q_{\pm} = \frac{1}{2} \, \left(u \, Q \,u^\dagger \pm u^\dagger \, Q \,
u\right) \,\, ,
\eeq
which under chiral SU(2)$_L\times$SU(2)$_R$ symmetry transform as
any matrix--valued matter field ($Q$ defines the charge matrix).

In particular, to lowest order  one finds (in the relativistic and
the heavy fermion formulation)
\beq
{\cal L}_{\pi N}^{(1)} = \bar{\Psi} \,\biggl( i \gamma_\mu \cdot
\tilde{\nabla}^\mu -
m + \frac{1}{2}g_A \, \gamma^\mu \gamma_5 \cdot \tilde{u}_\mu \, \biggr) \,
\Psi \quad  = \quad \bar{N} \,\biggl( i v \cdot \tilde{\nabla} +
g_A \, S \cdot \tilde{u} \, \biggr) \, N + {\cal O}(\frac{1}{m})\,\,\, ,
\eeq
with
\beq \label{covder}
\tilde{\nabla}_\mu = \nabla_\mu -i \, Q_+ \, A_\mu \,\,\, , \,\,\,
\tilde{u}_\mu = u_\mu - 2 \, Q_- \, A_\mu \,\,\, ,
\eeq
and\footnote{We do not spell out the details of how to construct the
heavy nucleon EFT from its relativistic counterpart but refer the
reader to the extensive review~\cite{bkmrev}.}
\beq
\Psi (x) = \exp\{ i mv \cdot x \} \, ( N(x) + h(x) )~.
\eeq
Furthermore, $v_\mu$ denotes the nucleons' four--velocity,
$S_\mu$ the covariant spin--vector \`a la Pauli--Lubanski
and $g_A$ the axial--vector coupling  constant. 
These virtual photon effects can only come in via loop diagrams since
by definition a virtual photon can not be an asymptotic state. 

At second order, local contact terms with finite low--energy
constants (LECs) appear. We call these LECs $f_i$ for the heavy baryon approach
and $f_i '$ in the relativistic Lagrangian. As stated before, the
em Lagrangian is given entirely in terms of squares of $Q_{\pm}$ (and
their traceless companions),
\beq\label{L2}
{\cal L}_{\pi N, {\rm em}}^{(2)}
= \sum_{i=1}^3 \, F_\pi^2 \, f_i ' \,  \bar{\Psi}\, {\cal O}_i^{(2)} \,\Psi
= \sum_{i=1}^3 \, F_\pi^2 \, f_i \,  \bar{N} \,{\cal O}_i^{(2)}\, N~,
\eeq
with the ${\cal O}_i^{(2)}$ monomials of dimension two,
\beq
{\cal O}_1^{(2)}  = \langle \qp^2 - \qm^2 \rangle~, \quad
{\cal O}_2^{(2)}  = \langle Q_+ \rangle \, \qp~, \quad
{\cal O}_3^{(2)}  =  \langle \qp^2 + \qm^2 \rangle~.
\eeq
Notice furthermore that only the second term in eq.(\ref{L2}) has an
isovector piece and contributes to the neutron--proton mass
difference~\cite{ms}. The factor $F_\pi^2$ in eq.(\ref{L2}) ensures
that the em LECs have the same dimension as the corresponding strong
dimension two LECs. From the third order calculation of the
proton--neutron mass difference~\cite{ms} one deduces the value
for $f_2$, $f_2 = -(0.45\pm 0.19)\,$GeV$^{-1}$.

The Lagrangian to third order takes the form
\beq \label{L3em}
{\cal L}^{(3)}_{\pi N,{\rm em}}
= \sum_{i=1}^{12} F_\pi^2 \, g_i' \, \bar \Psi \,{\cal O}_i^{(3)} \, \Psi \,\,
= \sum_{i=1}^{12} F_\pi^2  \,g_i \, \bar N \,{\cal O}_i^{(3)} \, N \,\, ,
\eeq
with the ${\cal O}_i^{(3)}$ monomials in the fields of dimension three 
\cite{ms}\cite{ug} , 
in their
relativistic form and the heavy baryon counterparts. Again, for the
$g_i$ to have the same mass dimension as the $d_i$ of the strong sector
defined in ref.\cite{FMS}, we have multiplied them with a factor of
$F_\pi^2$. Thus the $g_i \, (g_i')$ scales as [mass$^{-2}$].
So the complete fourth order pion--nucleon Lagrangian with
virtual photons is given by \cite{ug}
\beqa \label{L4em}
{\cal L}^{(4)}_{\pi N,{\rm em}}
&=& \sum_{i=1}^{5} F_\pi^4 \, h_i' \,
\bar \Psi \,{\cal O}_i^{(e^4)} \, \Psi \,\,
+ \sum_{i=6}^{90}  F_\pi^2 \, h_i' \,
\bar \Psi \,{\cal O}_i^{(e^2 p^2)} \, \Psi \,\, 
\eeqa
with the ${\cal O}_i^{(4)}$ monomials in the fields of dimension four
\cite{ug}. To be consistent with the
scaling properties of the dimension two and three LECs, the $h_i$ are
multiplied with powers of $F_\pi^2$ such that the
first five LECs take dimension [mass$^{-3}$] while the others are
of dimension [mass$^{-1}$].
 
\subsection*{The scalar form factor}
\label{sec:sigma}

The scalar form factor of the nucleon is defined via
\beq
\langle N(p') |\,  m_u \bar u u + m_d \bar d d \,| N(p)\rangle
= \bar{u} (p') u(p) \, \sigma (t)~, \quad t = (p'-p)^2~,
\eeq
for a nucleon state $|N(p)\rangle$ of four--momentum $p$. At $t=0$, which
gives the much discussed pion--nucleon $\sigma$--term, one can relate
this matrix element to the so--called strangeness content of the
nucleon. A direct determination of the $\sigma$--term is not possible,
but rather one extends pion--nucleon scattering data into the unphysical
region and determines $\sigma_{\pi N} (t=2M_\pi^2)$, i.e. at the so--called
Cheng--Dashen point. The relation to the $\sigma$--term is given by
the low--energy theorem of Brown, Peccei and Pardee~\cite{BPP},
\beq
\sigma_{\pi N}(2M_\pi^2) = \sigma_{\pi N}(0) + \Delta \sigma_{\pi N}
+ \Delta R
\eeq
where $\Delta \sigma_{\pi N}$ parametrizes the $t$--dependence of the
sigma--term whereas $\Delta R$ is a remainder not fixed by chiral symmetry.
The most systematic determination of $\Delta \sigma_{\pi N}$ has been
given in ref.\cite{GLS}, $\Delta \sigma_{\pi N}
= (15\pm 1)$~MeV. The remainder $\Delta R$ has been bounded in
ref.\cite{BKMcd},
$\Delta R < 2$~MeV. It was shown that the third order
effects can shift the proton $\sigma$--term by about 8\% and have a smaller
influence on the shift to the Cheng--Dashen (CD) point \cite{ms}. 
In \cite{ug} we worked out
explicitly the isospin violating corrections to this shift to fourth
order. This is motivated by the fact that in the difference most of the
counterterm
contributions drop out, more precisely, only momentum--dependent contact terms
can contribute to the shift. Such terms only appear at fourth order since
due to parity one needs two derivatives and any quark mass or em insertion
accounts for at least two orders \cite{ug}. 
It can be decomposed as
\beq
\sigma_{\pi N}^{(4)} (t) =  \sigma_{\pi N}^{\rm (4), IC} (t) +
\sigma_{\pi N}^{\rm (4), IV} (t)~.
\eeq
The isospin--conserving strong terms have already been evaluated in
ref.\cite{BKMcd}. Here, we concentrate on the em corrections $\sim
1$ (in isospin space) and all terms $\sim \tau_3$.
We have eye graphs and tadpoles with insertions $\sim f_2, c_5$. These
can be evaluated straightforwardly. Because
of the tiny coefficients appearing in the evaluation of the loop
contributions, these are only fractions of an
MeV, $\Delta \sigma_{\pi N}^{4, IV, \rm loop} = -0.05\,$MeV and can thus
be completely neglected. For the counterterm contributions, setting
all appearing LECs on the values obtained from dimensional
analysis as explained \cite{ug}, one
finds a total contribution $\Delta \sigma_{\pi N}^{4, IV, \rm ct} =
\pm0.01\,$MeV. For the IC em terms, we find
(setting again $f_{1,3} = \pm 1/4\pi$) a completely negligible loop
contribution (less than 0.01~MeV) and the counterterms give
$\pm 0.7\,$MeV for the LECs estimated from dimensional analysis.
Note, however, that if the numerical factors ${f}_{1,3}/(4\pi)$ are
somewhat bigger than one, one could easily have a shift of $\pm
2\,$MeV, which is a substantial electromagnetic effect.

\subsection*{Neutral pion scattering off nucleons}
\label{sec:pi0N}

As pointed out long time ago by Weinberg~\cite{weinmass}, the difference
in the S--wave scattering lengths for neutral pions off nucleons is
sensitive to the light quark mass difference,
\beqa\label{weinf}
a(\pi^0 p)-a(\pi^0 n) &=& \frac{1}{4\pi}\frac{1}{1+M_{\pi^+}/m_p} \,
\frac{-4B(m_u-m_d)c_5}{F_\pi^2} +{\cal O}(q^3) \nonumber \\
&=& \frac{1}{4\pi}\frac{1}{1+M_{\pi^+}/m_p} \,
\Delta_2 (\pn)  +{\cal O}(q^3)~.
\eeqa
It was shown in ref.\cite{ms} by an explicit calculation that to third order
there are no corrections to this formula. This is based on the fact that
the electromagnetic Lagrangian can not contribute at this order
since the charge matrix has to appear quadratic and never two
additional pions can appear. However, at next order one can of course have
loop graphs with one dimension two insertion and additional em
counterterms. To obtain the first correction to Weinberg's prediction,
eq.(\ref{weinf}), one thus has to compute the fourth order
corrections \cite{ug}.
These are due to strong dimension two insertions $\sim c_5$ and em insertions
$\sim f_2$. For the difference $a(\pi^0 p)-a(\pi^0
n)$
we only have to consider the operators $\sim \tau^3$ \cite{ug} 
Consider first the loop contributions. Since we can not fix the counterterms
from data, we are left with a spurious scale dependence which reflects the
theoretical uncertainty at this order. For $\lambda = \{0.5, 0.77, 1.0\}\,$GeV
we find
\beq
\Delta_2^{\rm str} = \{ -7.1 , 0.9 , 5.7 \} \cdot 10^{-2}~, \quad
\Delta_2^{\rm em}  = \{ 11.5 , 12.0 , 12.3 \}\cdot 10^{-2} ~.
\eeq
The counterterms are estimated based on dimensional analysis at the
scale $\lambda = M_\rho$ and give a contribution of about $-0.3
\cdot 10^{-2}$. Even if the LECs
would be a factor of ten larger than assumed, the counterterm contribution
would not exceed $\pm 3\%$.
 Altogether, the correction to Weinberg's prediction, eq.(\ref{weinf}),
are in the range of $4$ to $18$ percent, i.e. fairly small.
 Finally, we wish to mention that in ref.\cite{AMB}
isospin--violation for neutral pion photoproduction off nucleons
was discussed which allows one to eventually measure directly
the very small $\pi^0 p$ scattering length by use of the final--state
theorem.

\subsection*{Summary}
We have developed the whole mechanism to calculate isospin violation effects
in the framework of CHPT. In order to get the correct size
of isospin violation one has to include all non-isospin symmetric sources
like the electromagnetic interaction, the quark mass difference 
($\pi_0 - \eta$ mixing), explicit photon loops. In future calculations
one has to pin down the two LECs of the second order em Lagrangian $f_{1,3}$
which are not well known and are estimated only by dimensional arguments.  

\bigskip

\noindent {\bf Acknowledgments}

\noindent I am grateful to Ulf-G. Mei{\ss}ner for enjoyable collaboration. 

\vspace{1cm}



\bigskip


\begin{thebibliography}{99}
\frenchspacing
\bibitem{weinmass} S. Weinberg, Trans. N.Y. Acad. of Sci. 38 (1977) 185.\vs
\bibitem{ms} Ulf-G. Mei{\ss}ner and S. Steininger,
Phys. Lett.  B419 (1998) 403.\vs
\bibitem{ug} Ulf-G. Mei{\ss}ner and G.\,M\"uller, Nucl.Phys. B556 (1998) 
in press. \vs
\bibitem{FMSi} N. Fettes,  Ulf-G. Mei{\ss}ner and S. Steininger,
{\tt hep-ph/9811366}, Phys. Lett. B (in press).\vs
\bibitem{jm} E. Jenkins and A.V. Manohar, Phys. Lett. B255 (1991) 558.\vs
\bibitem{bkkm}V. Bernard, N. Kaiser, J. Kambor and Ulf-G. Mei\ss ner,
Nucl. Phys. B388 (1992) 315.\vs
\bibitem{gss} J. Gasser, M.E. Sainio and A. ${\rm \check S}$varc,
Nucl. Phys. B307 (1988) 779.\vs
\bibitem{bkmff} V. Bernard, N. Kaiser and Ulf-G. Mei{\ss}ner,
Nucl. Phys.  A611 (1996) 429.\vs
\bibitem{BL} T. Becher and H. Leutwyler, {\tt hep-ph/9901384}.\vs
\bibitem{FMS} N. Fettes,  Ulf-G. Mei{\ss}ner and S. Steininger,
Nucl. Phys. A640 (1998) 199.\vs
\bibitem{bkmrev} V. Bernard, N. Kaiser and Ulf-G. Mei{\ss}ner,
 Int. J. Mod. Phys.  E4 (1995) 193.\vs
\bibitem{MMS} Ulf-G. Mei{\ss}ner, G. M\"uller and S. Steininger,
{\tt hep-ph/9809446},  Ann. Phys.~(NY), in press.\vs
\bibitem{BPP} L.S. Brown, W.J. Pardee and R.D. Peccei, Phys. Rev. D4 (1971)
2801.\vs
\bibitem{GLS} J. Gasser. H. Leutwyler and M.E. Sainio, Phys. Lett. B253 (1991)
252.\vs
\bibitem{BKMcd}  V. Bernard, N. Kaiser and Ulf-G. Mei{\ss}ner, Phys. Lett.
B389 (1996) 144.\vs
\bibitem{AMB} A.M. Bernstein, Phys. Lett. B442 (1998) 20.\vs
\end{thebibliography}
\end{document}